# Violation of Bell inequalities by stochastic simulations of Gaussian States based on their positive Wigner representation

Eric Lantz[*], Mehdi Mabed, Fabrice Devaux

Institut FEMTO-ST, Université de Bourgogne Franche-Comté CNRS,

15 B rue des Montboucons 25030 Besançon, France

[*] e-mail : eric.lantz@univ-fcomte.fr

**Abstract**

At first sight, the use of an everywhere positive Wigner function as a probability density to perform stochastic simulations in quantum optics seems equivalent to the introduction of local hidden variables, thus preventing any violation of Bell inequalities. However, because of the difference between symmetrically and normally ordered operators, some trajectories in stochastic simulations can imply negative intensities, despite a positive mean. Hence, Bell inequalities do not apply. Here, we retrieve for a weakly squeezed Gaussian state the maximum violation on polarization states allowed by quantum mechanics, for the Clauser-Horn-Shimony-Holt (CHSH), as well as for the Clauser-Horn Bell inequalities. For the case of the Clauser-Horn Bell inequality, the influence of the quantum efficiency of the detectors is studied, and for both inequalities, the influence of the degree of squeezing is assessed, as well as the uncertainty range versus the number of trajectories used in the simulations.

## 1) Introduction

As reported by Drummond et al. [1], in 1982 Richard Feynman answered in the negative [2] to the question "Can quantum systems be probabilistically simulated by a classical computer?" Following [1] and others, we propose in this paper a more positive answer. However, we first would like to remark that there are simple systems that justify Feynman assertion. Consider for example two entangled images formed by spontaneous parametric down conversion (SPDC) measured with two cameras able to count photons [3]. Let the number of photons incident on each pixel of the cameras be smaller than, say, 5, and the number of relevant pixels on each camera be $512 \times 512$. In the Schrödinger point of view, the Hilbert space that describes such a system has $5^{2 \times 512 \times 512}$ dimensions, each corresponding to a different pair of images with its own probability. It is clearly impossible to simulate, by using a generator of random numbers corresponding to these probabilities, the successive pairs of images obtained by repeating the experiment. On the other hand, all statistical features of the images, like means, variances, pixel correlations and so on, can easily be obtained in the Heisenberg point of view, since quantum quadratures propagate like classical optical fields. To calculate the statistical features of the spatial repartition of SPDC, two methods were proposed in [4], using either Green's functions to characterize the pixel to pixel input-output relations, or stochastic simulations based on the Wigner phase-space representation.

Bell inequalities involve correlations between remote systems, and the demonstration of their violation can be performed using the characteristic function, i.e. the Fourier transform of the Wigner function [5]. Besides, Bell argued [6] that the Wigner function of a pair of position-momentum entangled particles, in the sense of Einstein-Podolsky-Rosen [7], can be seen as a probability distribution for position and momentum of a pair of classical particles, preventing any non-locality. Though, as mentioned explicitly in [6], the argument did not apply to dichotomic variables like spin or polarization, the discussion that followed in the subsequent literature mentioned important elements concerning the possibility of using the Wigner function as a probability density for stochastic simulations. Notably, Ref. [8] stated that the Wigner representation of quantum observables cannot be in general interpreted as phase-space distribution of possible experimental outcomes. For spatial variables, a relation between the Wigner function and the parity operator was used in [8] to show violation of Bell inequalities, as experimentally demonstrated in spatial parity space using SPDC [9]. Note that SPDC, or squeezed vacuum, does possess a positive-definite Wigner function [10], which leads to Bell inequalities only for variables having a definite value assigned by the local hidden variables [11].

Actually, before these works in the spatial domain, SPDC was proved by Kwiatt et al to allow a violation of the Bell inequalities in the original scheme using polarizers [12]. The experiment used polarization entanglement between the pairs of photons coming from the two intersections of the two cones corresponding to type-II SPDC, each cone with a polarization orthogonal to the other: see Figure 2. At the point of detection, the polarizing beam-splitters can be viewed as parity operators. This experiment was analyzed in the Wigner representation by Casado et al [13], who remarked that the detected intensity is not positive-definite for each realization of the underlying random process, even if the mean intensity, corresponding in simulations to an average of a large number of realizations, is positive. This remark led them to propose a modification of the quantum formalism that is compatible with experiment only if the detectors admit some basic dark rate, or in other words, if they are directly sensitive to vacuum fluctuations. No experimental evidence has justified this proposition, beyond the evident problems of energy conservation. Brambilla et al. used also stochastic simulations based on the Wigner formalism to describe the spatial properties of SPDC [14].

A positive phase-space Wigner function offers a /straightforward/clear scheme for stochastic simulations that should, for polarization entangled SPDC, violate Bell inequalities since, in the words of Cahill and Glauber [15], the Wigner function may be used to write the ensemble averages of all bounded operators as convergent integrals. We will see in this paper that this is indeed the case. Curiously, such violation seems have not been demonstrated before, though Werner et al presented [16] a thorough comparison of the advantages and drawbacks of the positive-P and Wigner representations. For our present purpose, we retain that the symmetrically ordered operators corresponding to the electric field in the Wigner representation (i.e. the quantum quadratures) propagate classically and the results are exact inasmuch as the pump field is intense and undepleted. These features were exploited in [17] to simulate a spatially multimode Hong-Ou-Mandel experiment, and in [18] to characterize SPDC issued from crystals with complex structures. As also stated in [16], the Wigner representation requires only half the number of variables as does the positive-P method, and requires independent Gaussian noise sources only at the input. Nevertheless, this is the positive-P method that was chosen to show that quantum simulations can be used to demonstrate violation of Bell inequalities [19]. As regards the Wigner formalism, it implies for each trajectory four complex numbers, corresponding to the two orthogonal polarizations of the field at the two remote locations. It has been demonstrated in [20-21] that four complex numbers defined in a similar way obey the Bell inequalities in the Glauber-Sundarshan representation. A misconception would consist in believing that this demonstration extends to the Wigner representation. Indeed, it is well-known [15, 20] that squeezed vacuum does not possess a regular Glauber-Sundarshan representation, though its Wigner function is positive-definite. On the other hand, the demonstration of Bell inequalities involves positive intensities. Hence Bell inequalities can be violated for a squeezed vacuum either because its Glauber-Sundarshan representation is not regular, or because the real number giving the intensity associated with a particular trajectory in the Wigner representation may be negative, as we will develop in this paper. Of course, it means that a trajectory of the stochastic simulation does not correspond to a possible experimental outcome [8] (looking for such a correspondence leads to doubtful physics [13]). Only averages computed over a great number of trajectories correspond to physical quantities. Nevertheless, the results for each trajectory are obtained by simulating classical propagation, ensuring that all details of the actual experiment can be easily taken into account [4, 12, 18].

The remainder of the paper is organized as follows. In section 2, we give the theoretical framework. Section 3 deals with numerical results and we conclude in section 4.

**2) Theoretical framework**

It has been shown by Cahill and Glauber, [15] Eq. 4.23, that the expectation value of a symmetrically ordered product of creation and annihilation operators $a^\dagger$ and $a$, can be always expressed as an integral in the entire complex plane C over a c-number $\alpha$, weighted by the Wigner function $W(\alpha)$:

$$< (a^\dagger)^n a^m >_S = \frac{1}{\pi} \int_C W(\alpha) \, (\alpha^*)^n \alpha^m d^2\alpha \qquad (1)$$

The subscript S, or symmetrically ordered, means that all orders are present with an equal weight in the expectation. For example, we have, for n=m=2:

$$< (a^\dagger)^2 a^2 >_S = \langle \frac{a^\dagger a^\dagger a a + a^\dagger a a^\dagger a + a^\dagger a a a^\dagger + a a^\dagger a a + a a^\dagger a a^\dagger + a a a^\dagger a^\dagger}{6} \rangle \qquad (2)$$

Some useful relations hold between the operator number of photons $N = a^\dagger a$, and the symmetrically ordered operators:

$$N = (a^\dagger a)_S - \frac{1}{2}, \qquad N^2 = ((a^\dagger)^2 a^2)_S - N - \frac{1}{2} \qquad (3)$$

Eqs. 3 are derived by using the commutation relation of the annihilation operator $[a, a^\dagger] = 1$.

We deduce the mean and the variance:

$$< N > = \langle (a^\dagger a)_S \rangle - \frac{1}{2}, \; V(N) = < N^2 > - < N >^2 = < (a^\dagger)^2 a^2 >_S - \langle (a^\dagger a)_S \rangle^2 - \frac{1}{4} \qquad (4)$$

Since these relations are based only of the commutation properties of the annihilation operator in a mode, they are general, whatever the wave function involved in the means.
If two different modes are implied, the corresponding annihilation operators commute, and we obtain for the covariance of the numbers of photons in two modes 1 and 2:

$$Cov(N_1, N_2) = < N_1 N_2 > - \langle N_1 \rangle \langle N_2 \rangle = < a_1^\dagger a_1 a_2^\dagger a_2 >_S - \langle (a_1^\dagger a_1)_S \rangle \langle (a_2^\dagger a_2)_S \rangle \qquad (5)$$

Eq. (1, 4, 5) suggests a scheme of numerical simulation for states whose Wigner function is positive definite, and remains as such under propagation. To calculate the integral on the right side of Eq. 1, the simplest solution is to randomly sample the complex plane by using a probability density proportional to the Wigner function. The real part of the obtained c-number corresponds to the position quadrature, quadrature $X_1 = \frac{a^\dagger + a}{2}$ in optics, and the imaginary part to the momentum quadrature, quadrature $X_2 = i\frac{a^\dagger - a}{2}$ in optics. It can be easily verified that $X_1^2 + X_2^2 = (a^\dagger a)_S$. Eq.1 ensures that the quantum mean of $X_1^2 + X_2^2$ is equal to the average of the squared moduli of the numerous randomly-drawn complex numbers. Clearly, the equality does not hold for an individual draw: if acting on the vacuum, $X_1$ and $X_2$ have a negative covariance, since their commutator is equal to $\frac{i}{2}$, while the real and imaginary parts of the c-numbers are independent. Indeed, the Wigner function of the vacuum is Gaussian and depends on the squared modulus, (see [15] Eq. 4.38):

$$W_0(\alpha) = 2\exp(-2|\alpha|^2) \tag{6}$$

We have now all the elements to introduce a complete numerical scheme to model the SPDC in realistic conditions corresponding to the different experimental set-up described in [4,12,18].

- Divide the input plane of the crystal in sufficiently small pixels to ensure that the sampling theorem is fulfilled at the crystal output. Indeed, phase-matching acts in the spatial domain as a low-pass amplifier, ensuring a spatial cutoff-frequency that defines the conditions of sampling. The sampling spatial frequency must be greater than twice the highest spatial frequency for which phase matching allows a non-negligible gain. As an example, we see on fig. 2 that the rings due to phase matching are entirely retrieved.
- Draw at random for each pixel two c-numbers, whose real and imaginary parts are independently selected from a Gaussian distribution of zero mean and variance ¼, in accordance with Eq. 6. Each c-number correspond to a polarized field along one of the two neutral axes of the crystal, horizontal (H) or vertical (V). We can prove easily from Eqs. (1, 4, 5), that such a draw ensures $<N> = 0$, $V(N) = 0$, Cov(N1,N2)=0 (two pixels 1 and 2), as expected for the input vacuum.
- Propagate the field in the crystal using the usual split-step algorithm, where the classical coupled equations of parametric amplification are solved in the direct domain, and diffraction is taken into account by propagating the plane wave spectrum in the spatial Fourier domain [4]. It can be proved that quantum quadratures propagate like classical waves in the undepleted pump approximation [16].
- Repeat the entire procedure many times. Each iteration is called a trajectory.
- Calculate at the output all the statistical features of interest on the detected photon-numbers by applying first Eq. 1, to pass from averages of squared moduli of c-numbers to means of symmetrically operators, then Eqs 3 to 5 to apply "quantum corrections" in order to retrieve photon numbers from symmetrically ordered operators.
  Note that these quantum corrections can be applied either to each trajectory (Eq.3) or to the means (Eqs 4 and 5).

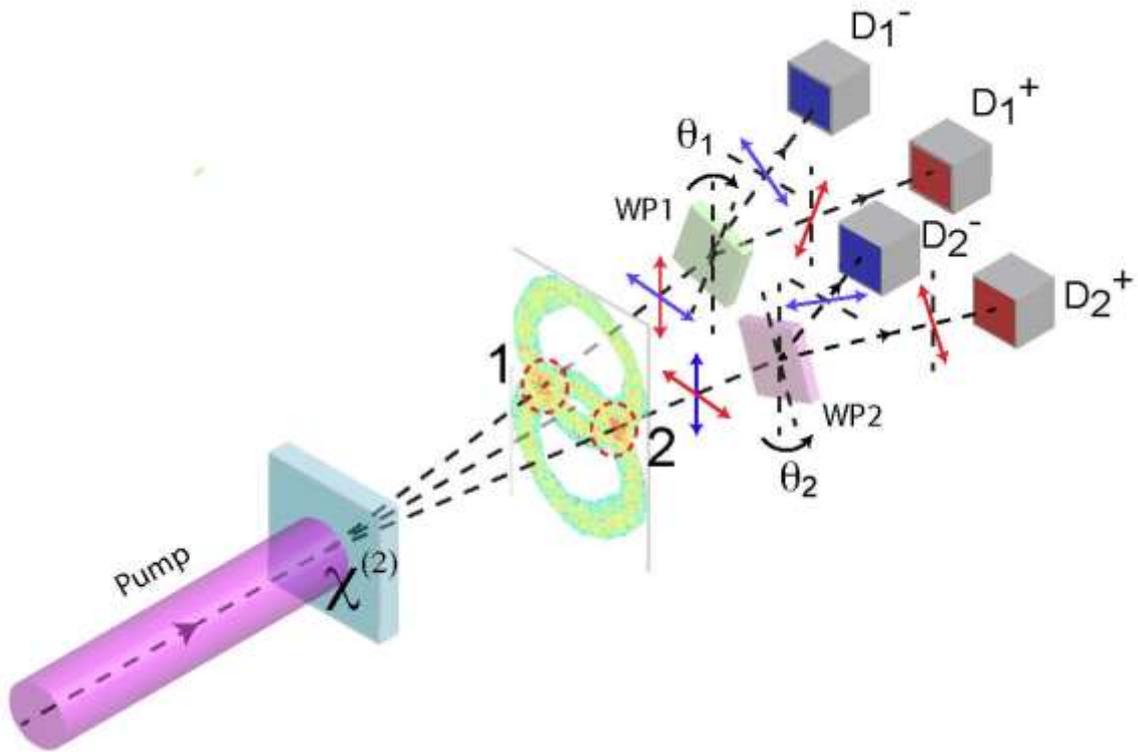

Fig.1: Experimental set-up. $\chi^{(2)}$: non linear crystal. WP: polarizing beam-splitters. D: single-photon detectors.

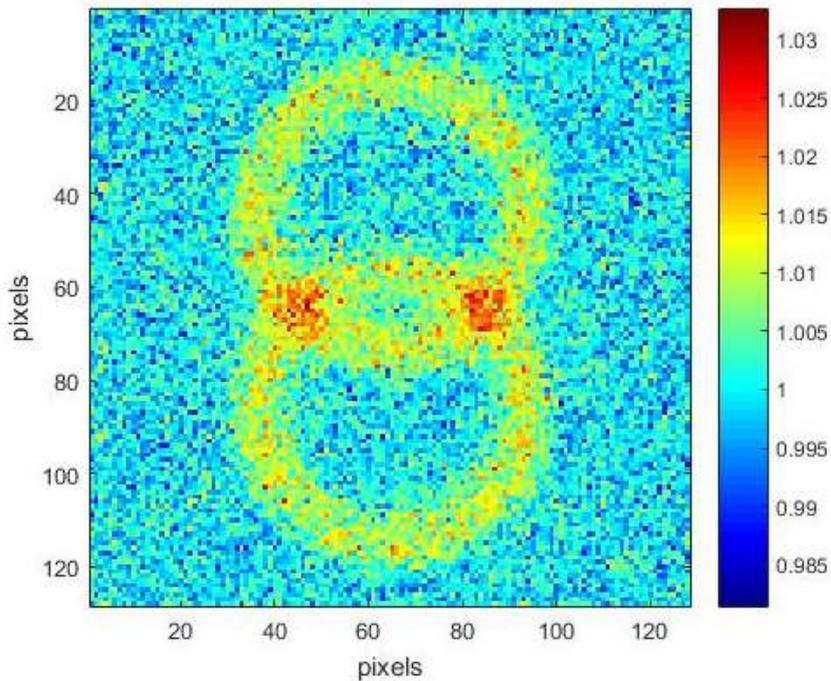

Fig2: Non corrected mean intensity in the Fourier plane (average of 20,000 trajectories).

The proposed experimental set-up is similar to that of [12]: see Fig.1. An U.V. pump beam is incident on a BBO crystal in conditions of type II phase-matching. A horizontally polarized signal and a vertically polarized idler beam are created by SPDC and four detectors record the photons coming from the cone intersections, in chosen polarization directions. We see on Fig. 2 the mean intensity obtained in the far-field, or Fourier domain, by averaging 30 000 trajectories. We keep for the following only two pixels, numbered 1 and 2, corresponding respectively to the left and right best intersection of the cones, exactly symmetrical with respect to the direction of the pump beam. The Bell biphoton state corresponding to these two pixels can be written as:

$$|\psi^+> = (|H1, V2> + |V1, H2>)/\sqrt{2} \qquad (7)$$

The other Bell states can be obtained by using wave-plates [12] but, for sake of conciseness, we consider only $|\psi^+>$ in this paper.

Two polarizing beam-splitters separate the beams 1 and 2, with their first neutral axes forming respectively an angle $\theta_1$ and $\theta_2$ with the horizontal direction. The four output field amplitudes can be written as:

$$\begin{pmatrix} A_i^+ \\ A_i^- \end{pmatrix} = \begin{pmatrix} \cos(\theta_i) & \sin(\theta_i) \\ -\sin(\theta_i) & \cos(\theta_i) \end{pmatrix} \begin{pmatrix} A_i^H \\ A_i^V \end{pmatrix} \qquad (8)$$

where + and – designate the two output ports of the polarizing beam-splitter and i=1 or 2 refers to the left or right pixel.

After calculating the intensities $I_i^j(\theta_i) = |A_i^j|^2$ (expressed in number of photons), we apply the quantum corrections of Eq.3 to obtain the normalized correlation:

$$E(\theta_1, \theta_2) = \frac{\overline{\left((I_1^+(\theta_1)-1/2)-(I_1^-(\theta_1)-1/2)\right)\left((I_2^+(\theta_2)-1/2)-(I_2^-(\theta_2)-1/2)\right)}}{\overline{\left((I_1^+(\theta_1)-1/2)+(I_1^-(\theta_1)-1/2)\right)\left((I_2^+(\theta_2)-1/2)+(I_2^-(\theta_2)-1/2)\right)}} \qquad (9)$$

The bars describe the average over a large number of trajectories n, n typically in the range $10^5$-$10^6$, as discussed in the next section. Clearly, the quantum corrections vanish in the numerator of Eq. (9). On the other hand, these quantum corrections do hold in the denominator. If we assume that the corrected intensities, i.e. the photon numbers, are positive for each trajectory, we can derive [20] the CHSH form of Bell inequalities [22]: whatever the angles $\theta_1, \theta_1', \theta_2, \theta_2'$, we have

$$|B| \leq 2, \text{ where } B = E(\theta_1, \theta_2) - E(\theta_1, \theta_2') + E(\theta_1', \theta_2') + E(\theta_1', \theta_2) \qquad (10)$$

However, for some trajectories, the photon numbers are negative because of the quantum corrections, although the average is (and must be) positive. Hence, the CHSH Bell inequality can be violated in our numerical experiment.

The experiment corresponding to the CHSH equality involves only measurements of coincidences and is therefore subject to the so called "fair sampling loophole". To avoid this loophole, Clauser and Horne proposed [23] an inequality involving the probability of detection of a single photon. For our purpose, this inequality can be written as:

$$C \leq 1, where\ C = \frac{\overline{I_1^+(\theta_1)I_1^+(\theta_2) - I_1^+(\theta_1)I_1^+(\theta_2') + I_1^+(\theta_1')I_1^+(\theta_2') + I_1^+(\theta_1')I_1^+(\theta_2)}}{\overline{I_1^+(\theta_1') + I_1^+(\theta_2)}} \qquad (11)$$

Since the denominator involves probabilities of detection of single photons, proportional to the detector quantum efficiency, while the numerator involves probabilities of coincidences proportional to the square of the efficiency, this inequality can be violated only for a high detector quantum efficiency, in fact greater than 83% [24] for a maximally entangled state. Once more, we will see that the negative "corrected intensities" allow this inequality to be numerically violated.

**3) Results**

All results will be given for $\theta_1 = -\frac{\pi}{8}$, $\theta_1' = \frac{\pi}{8}$, $\theta_2 = \frac{\pi}{2}$, $\theta_2' = -\frac{\pi}{4}$, ensuring for the Bell state $|\psi^+\rangle$ the quantum theoretical values $B = 2\sqrt{2} = 2.83$, $C = \frac{1+\sqrt{2}}{2} = 1.21$, i.e. the maximum violation of the Bell inequalities allowed by quantum mechanics. See for example [20] for the quantum calculation.

Fig. 3 shows, for a mean output intensity of 0.02 photon/pixel (sum of signal and idler), the evolution of B and C for a number of trajectories n between 1000 and $1400^2 = 1.96\ 10^6$. For the 128x128 pixels considered in this image, this simulation corresponds to 7 hours of computation time on a professional PC. A simple estimation of the confidence interval for the intensities is as follows. The probability density of the non-corrected intensity for a mode, i.e. after the polarizing beam-splitter, is given by a decreasing exponential, with, for a trajectory, a standard deviation equal to the mean (the statistics is that of a speckle of unity contrast [25]). The standard deviation of the total mean intensity $\sigma_{\bar{I}}$ is inversely proportional to the square root of the number of trajectories, giving for $n=1.96\ 10^6$ and a mean non-corrected intensity of the signal or the idler $\bar{I}_{sNC} = 0.51$, $\sigma_{\bar{I}} = \frac{0.51 \times \sqrt{2}}{1400} = 5.2\ 10^{-4}$. We use here the fact that the signal and the idler intensities have independent statistics when adding on either the pixel 1 or 2 and each obey thermal statistics (standard deviation equal to the mean, see below). On the other hand, the true intensity is the corrected one, giving a relative standard deviation of $\frac{5.2\ 10^{-4}}{2\ 10^{-2}} = 2.6\%$. Though the exact computation of the uncertainty range for B is difficult, we can admit that this value is also close to the relative standard deviation of B. We see here the principal drawback of the method: the useful information lies in the corrected values, while the fluctuations scale with the non-corrected ones, leading to the necessity of a great number of trajectories for the small gain that allows a weak squeezed state to reproduce at best the quantum behavior of a biphoton state.

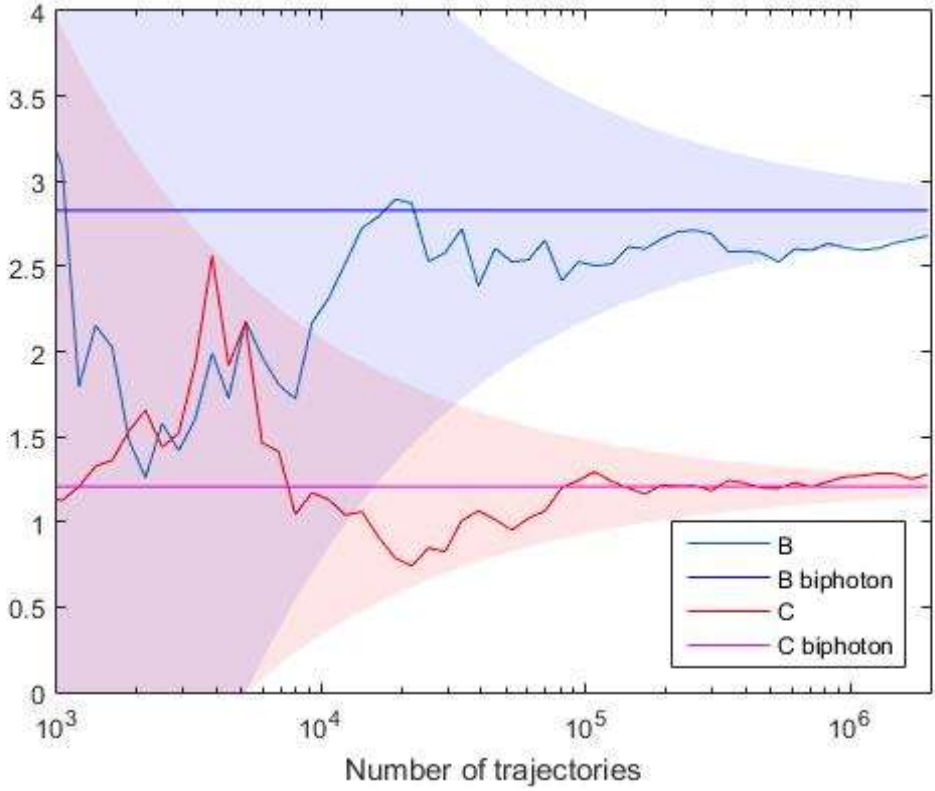

Fig.3: An example of the evolution of B and C versus the number of trajectories. Colored areas: uncertainty ranges (95% of confidence) centered on the theoretical biphoton values.

The finally estimated B=2.68 is 5.3% below the quantum theoretical value for a biphoton state, B=2.83, i.e. outside the $\pm 5.2\%$ uncertainty range at 95% confidence. Actually, even for a low mean intensity of 0.02, the probability of a double pair in a single experiment cannot be entirely neglected. It leads to a modification of the coincidence rate that lowers the measured value of B. The theoretical value of B that takes into account this effect can be determined as follows. Let be $G = sinh^2(g\,L)$ the total gain, in photons per mode, for a crystal of length L. $g$ is the gain per unit length, depending on the pump intensity and the nonlinear crystal coefficient, at perfect phase matching. The statistics of the signal (idler) beam is thermal, ensuring for its mean and variance [25]:

$$\langle N_S \rangle = \langle N_I \rangle = G, \quad V(N_S) = V(N_I) = G + G^2 \qquad (12)$$

Only pairs are emitted, resulting in a signal-idler covariance equal to the variance:

$$Cov(N_S, N_I) = G + G^2 \qquad (13)$$

At the intersection of the cones, the signal and idler intensities are added and not correlated, ensuring:
$$\langle N_1 \rangle = \langle N_2 \rangle = 2G, \quad V(N_1) = V(N_2) = 2\,(G + G^2) \qquad (14)$$

There is perfect correlation between the signal (idler) in 1 and the idler (signal) in 2, which allows us to write the covariance between the two pixels as:

$$Cov(N_1, N_2) = 2Cov(N_S, N_I) = 2(G + G^2) \tag{15}$$

We now find easily all terms necessary to compute the theoretical value of $E(\theta_1, \theta_2)$:

$$\langle N_1 N_2 \rangle = 2G + 6G^2$$
$$\langle (N_1^+(\theta_1) - N_1^-(\theta_1))(N_2^+(\theta_2) - N_2^-(\theta_2)) \rangle = 2(G + G^2)(sin^2(\theta_1 + \theta_2) - cos^2(\theta_1 + \theta_2)) \tag{16}$$

Giving, for the angles corresponding to a maximum violation:

$$B = 2\sqrt{2}\left(\frac{1+G}{1+3G}\right) \tag{17}$$

For G=0.01 used in Fig. 2, the theoretical value of B is 2.77, i.e. well inside the $\pm 5.2\%$ uncertainty range around the "experimental" value 2.68.

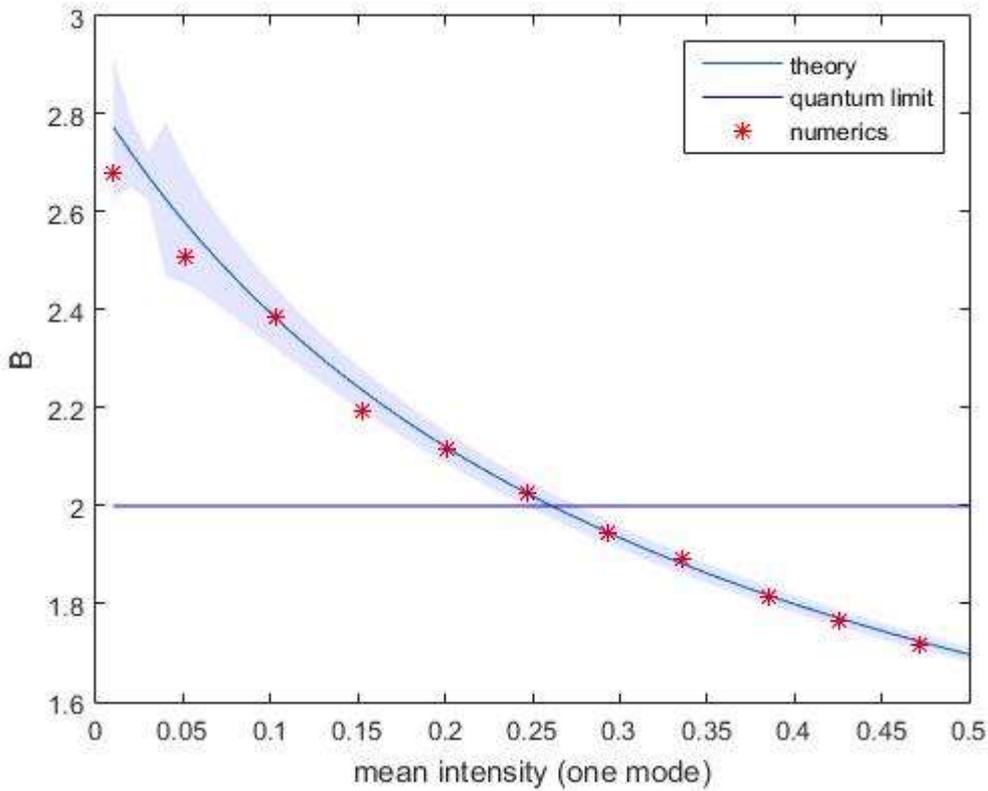

Figure 4: Numerical and analytical (eq.17) values of B versus the intensity G in a mode. Colored area: uncertainty range.

Figure 4 shows the comparison between the values of B issued from the numerical simulation ($10^5$ trajectories for all points but the first, 1.96 x $10^6$ trajectories for this point) and the values calculated with Eq. 16, with a good agreement. It is also interesting to note that the relative number of negative values of $N_1 N_2$ goes from 47% for G=0.01 to 22% for G=0.46. The quantum limit B=2 is attained for 31% of negative values.

It should be noted that $C$ increases with G. We see immediately from the mean in Eq. 14 and the correlations in Eq. 16 that, for angles corresponding to a maximum violation and unity quantum efficiency, we have:

$$C = 1.21 \frac{2(G+G^2)}{2G} = 1.21(1+G) \qquad (18)$$

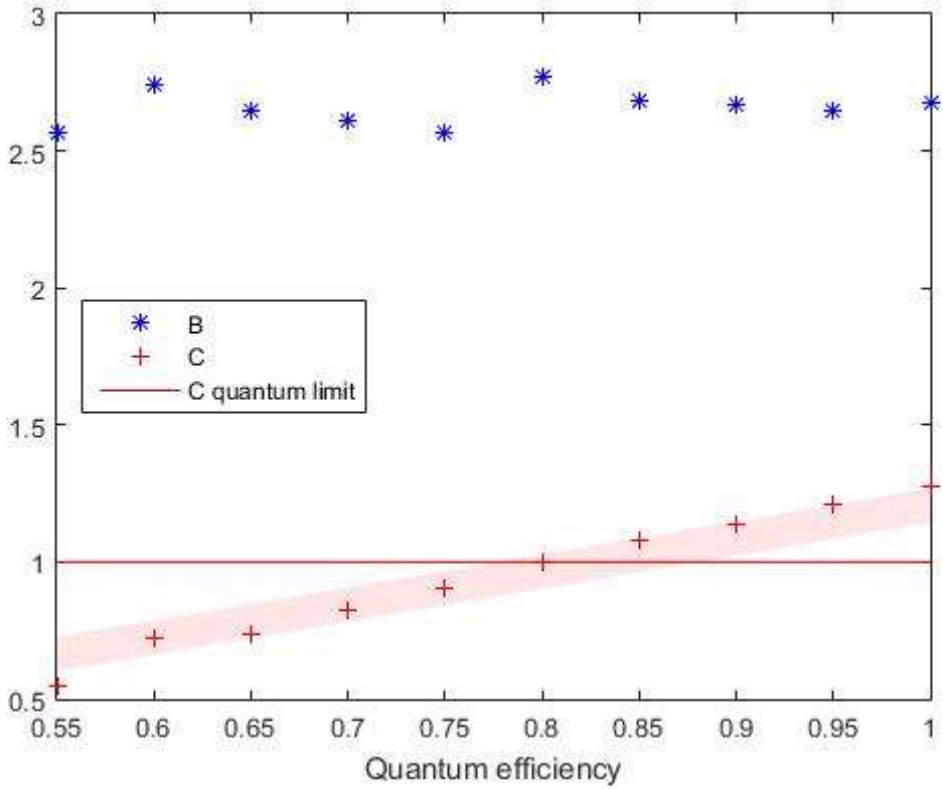

Figure 5: B and C versus the quantum efficiency, for an intensity per mode of 0.01. The colored uncertainty range is centered on the maximum value ($\frac{1+\sqrt{2}}{2} = 1.21$) multiplied by the quantum efficiency.

Finally we see in Fig.5 the influence of the quantum efficiency, equivalent to a beam splitter before each detector, with quantum vacuum noise entering the free input port. As foreseen, B is independent of the quantum efficiency, while C surpasses the quantum limit 1 only for a high quantum efficiency, since C is simply proportional to the quantum efficiency $\eta$:

$$C = \frac{1+\sqrt{2}}{2} \eta \qquad (19)$$

proving non-locality for a minimum quantum efficiency [24] $\eta = \frac{2}{1+\sqrt{2}} = 0.83$

**4) Conclusion**

We have shown in this paper that stochastic simulations based on the positive Wigner function of Gaussian states can be used to demonstrate violation of Bell inequalities. The method is simpler than the positive P representation [16], requiring for each trajectory four complex numbers instead of eight. The minimum of trajectories to attain a good precision becomes very important if the mean number of photons per mode is very low, i.e. in the regime where the probability of a second pair in the mode is weak, meaning that the simulation corresponds to a genuine biphoton. Nevertheless, strong violation (B=2.6) can be obtained with a mean number of photons per mode of 0.05 and $10^5$ trajectories, i.e. 20 minutes on a professional PC, and an uncertainty of about $\pm 4\%$.

These results allow the degree of nonlocality to be assessed for realistic experimental conditions, with a classical propagation simulation that can integrate all experimental details, for example a non-ideal pump shape as in [4], or a periodically poled crystal as in [18].

Of course, Bell inequalities are well known for polarization entanglement and, once validated in this case, our method could be employed to less explored situations. We envisage extending our method to high-dimensional systems [26, 27], where the reduced number of variables, compared to the positive-P, could be very interesting.

## Bibliography

[1] P D Drummond, B Opanchuk, L Rosales-Arate, and M D Reid. "Simulating Bell violations without quantum computers". Physica Scripta **160** 014009 (2014),

[2] R.P.Feynman, "Simulating physics with computers" Int. J. Theor. Phys. **21** 467–488 (1982)

[3] P.A Moreau, F. Devaux, and E. Lantz, "Einstein-Podolsky-Rosen Paradox in Twin Images", Phys. Rev. Letters **113,** 160401 (2014)

[4] E. Lantz, N. Treps, C. Fabre and E. Brambilla, " Spatial distribution of quantum fluctuations in spontaneous down-conversion in realistic situations: comparison between the stochastic approach and the Green's function method", Eur. Phys. Journal D **29** 437-444 (2004).

[5] Y. Tsujimoto, K. Wakui, M. Fujiwara  K. Hayasaka, S. Miki, H. Terai, M. Sasaki, and M. Takeoka, "Optimal conditions for the Bell test using spontaneous parametric down-conversion sources", Phys. Rev. A **98**, 063842 (2018)

[6] J. S. Bell, "EPR Correlations and EPW Distributions", in: New Techniques and Ideas in Quantum Measurement Theory, ed.D. M. Greenberger NY Acad. Sci. **480**, 263 (1986).

[7] A. Einstein, B. Podolsky, and N. Rosen, "Can quantum-mechanical description of physical reality be considered complete?" Phys. Rev. 47, 777–780 (1935)

[8] K. Banaszek and K. Wodkiewicz, "Nonlocality of the Einstein-Podolsky-Rosen state in the Wigner representation", Phys. Rev. A **58**, 4345 (1998)

[9] T. Yarnall, A. F. Abouraddy, B. E. A. Saleh and M. C. Teich, **"**Experimental Violation of Bell's Inequality in Spatial-Parity Space", Phys. Rev. Letters **99,** 170408 (2007)

[10] By taking into account the exact spatial phase matching profile, small negativities can appear, leading to a very weak violation of Bell  inequalities for position-momentum variables: J. Schneeloch, S. H. Knarr, D. J. Lum, and J. C. Howell, "Position-momentum Bell nonlocality with entangled photon pairs", Phys. Rev. A **93**, 012105 (2016)

[11] M. Revzen, P. A. Mello, A. Mann and L. M. Johansen, "Bell's inequality violation with non-negative Wigner functions", Phys. Rev. A **71**, 022103 (2005)

[12] P.G. Kwiatt, K. Mattle, H. Weinfurter and A. Zeilinger, "New High-Intensity Source of Polarization-Entangled Photon Pairs", Phys. Rev. Letters  **75**, 4337-4341 (1995)

[13] A. Casado, T. W. Marshall and E. Santos "Type II parametric down-conversion in the Wigner-function formalism: entanglement and Bell's inequalities", J. Opt. Soc. Am. B  **15**, 1572-1577 (1998)

[14] E. Brambilla, A. Gatti, M. Bache, and L. A. Lugiato, "Simultaneous near-field and far-field spatial quantum correlations in the high-gain regime of parametric down-conversion**",** Phys. Rev. A **69**, 023802 (2004)

[15] K. E. Cahill and R. J. Glauber, "Density Operators and Quasiprobability Distributions", Phys. Rev. **177** 1882-1902 (1969)

[16] M.J. Werner, M.G. Raymer, M. Beck and P. D. Drummond, "Ultrashort pulsed squeezing by optical parametric amplification", Phys. Rev. A **52** 4202-4212 (1995)

[17] F. Devaux, A. Mosset and E. Lantz," Stochastic numerical simulations of a fully spatiotemporal Hong-Ou-Mandel dip", Phys. Rev. A **100,** 013845 (2019)

[18] S.Trajtenberg-Mills, A. Karnieli, N. Voloch-Bloch, E. Megidish, H.S. Eisenberg, and A. Arie, "Simulating Correlations of Structured Spontaneously Down-Converted Photon Pairs", *Laser Photonics Rev.* **14**, 1900321 (2020)

[19] L. Rosales-Zarate, B. Opanchuk, P. D. Drummond, and M. D. Reid , **"**Probabilistic quantum phase-space simulation of Bell violations and their dynamical evolution", Phys. Rev. A **90**, 022109 (2014)

[20] D.F. Walls and G.J. Milburn, "Quantum Optics", 2nd edition, Springer (2008)

[21] M. D. Reid and D. F. Valls, "Violations of classical inequalities in quantum optics", Phys. Rev. A **34,** 1260-1276 (1986)

[22] J.F. Clauser, M.A. Horne, A. Shimony and R.A. Holt, "Proposed experiment to test local hidden-variable theories", Phys. Rev. Lett., **23** pp 880–884 (1969)

[23] ] J.F. Clauser and M.A. Horne , "Experimental consequences of objective local theories", Phys. Rev. D **10**, 526-535 (1974)

[24] A. Garg and N.D. Mermin. "Detector inefficiencies in the Einstein-Podolsky-Rosen experiment". Phys. Rev. D. **35** 3831–5 (1987)

[25] J. W. Goodman, "Statistical Optics", 2nd edition, Wiley (2015)

[26] D. Collins, N. Gisin, N. Linden, S. Massar, and S. Popescu, "Bell Inequalities for Arbitrarily High-Dimensional Systems", Phys. Rev. Lett., **88,** 040404 (2002)

[27] W. Li and S. Zhao**,** "Bell's inequality tests via correlated diffraction of high-dimensional position-entangled two-photon states", Scientific reports **8** 4812 (2018)